\documentstyle[emulateapj,onecolfloat,psfig]{article}

\newcommand{\lin}{{\rm pt}}
\newcommand{\shell}{{\rm s}}
\newcommand{\dirac}{{\rm D}}
\newcommand{\halo}{{\rm h}}

\newcommand{\bp}{{\cal C}} 

\newcommand{\bfx}{{\mathbf{x}}}
\newcommand{\bfl}{{\mathbf{l}}}

\newcommand{\veck}{{\bf k}}
\newcommand{\vecl}{{\bf l}}
\newcommand{\vecr}{{\bf r}}
\newcommand{\vecx}{{\bf x}}

\newlength{\tskip}
\newlength{\colwidth}

\newcommand{\beq}{\begin{equation}}
\newcommand{\eeq}{\end{equation}}
\newcommand{\beqa}{\begin{eqnarray}}
\newcommand{\eeqa}{\end{eqnarray}}

\def\simgt{\gtrsim}

\newcommand{\rad}{r}    

\newcommand{\sky}{{\rm sky}}

\begin{document}
\twocolumn[
\title{Power Spectrum Covariance of Weak Gravitational Lensing}
\author{Asantha Cooray and Wayne Hu}
\affil{
Department of Astronomy and Astrophysics, University of Chicago,
Chicago, IL 60637\\
E-mail: asante@hyde.uchicago.edu, whu@background.uchicago.edu}


\begin{abstract}
Weak gravitational lensing observations probe the spectrum
and evolution of density fluctuations and the cosmological parameters
which govern them. At low redshifts, the non-linear gravitational 
evolution of large scale structure produces a non-Gaussian covariance 
in the shear power spectrum measurements that affects their 
translation into cosmological  parameters.  Using the dark matter halo approach, 
we study the covariance of binned band power spectrum estimates and the four point function
of the dark matter density field that underlies it. 
We compare this semi-analytic estimate to results from N-body 
numerical simulations and find good agreement.
We find that for a survey out to z $\sim$ 1, the power spectrum covariance
increases the errors on cosmological parameters determined under the
Gaussian assumption by about 15\%. 
\end{abstract}

]

\section{Introduction}

Weak gravitational lensing 
by large scale structure (LSS) shears
the images of faint galaxies at the percent level
and correlates their measured ellipticities  
(e.g., \cite{Blaetal91} 1991; \cite{Mir91} 1991; 
\cite{Kai92} 1992).  Though challenging to measure, the two point
correlations, and the power spectrum that underlies them,  
provide important cosmological information that is
complementary to that supplied by
the cosmic microwave background and potentially as precise
(e.g., \cite{JaiSel97} 1997;
\cite{Beretal97} 1997; \cite{Kai98} 1998; 
\cite{HuTeg99} 1999; \cite{Hui99} 1999; \cite{Coo99} 1999; \cite{Vanetal99} 1999;
see \cite{BarSch00} 2000 for a recent review).
Indeed several recent studies have provided the first clear evidence
for weak lensing in so-called blank fields where the large scale
structure signal is expected to dominate (e.g., \cite{Vanetal00}
2000; \cite{Bacetal00} 2000; \cite{Witetal00} 2000; \cite{Kaietal00} 2000).
 
Given that weak gravitational lensing probes the projected mass
distribution, its statistical properties reflect those of the dark matter. 
Non-linearities in the mass distribution, due to gravitational
evolution at low redshifts, causes the shear field to become non-Gaussian.
It is well known that lensing induces a measureable three-point
correlation in the derived convergence field (\cite{Beretal97} 1997;
\cite{CooHu00} 2000).  The same processes also induce a four point correlation.
The four point correlations are of particular interest in that
they quantify the sample variance and covariance 
of two point correlation or power spectrum
measurements.  
Previous studies of the ability of 
power spectrum measurements to constrain cosmology
have been based on a Gaussian approximation to the sample variance
and the assumption that covariance is negligible
(e.g., \cite{HuTeg99} 1999); it is
of interest to test to what extent their inferences remain valid 
in the presence of realistic non-Gaussianity.  More importantly,
when interpreting the power spectrum recovered from the
next generation of surveys an accurate propagation of errors
will be critical (\cite{HuWhi00} 2000).

\begin{figure}[t]
\centerline{\psfig{file=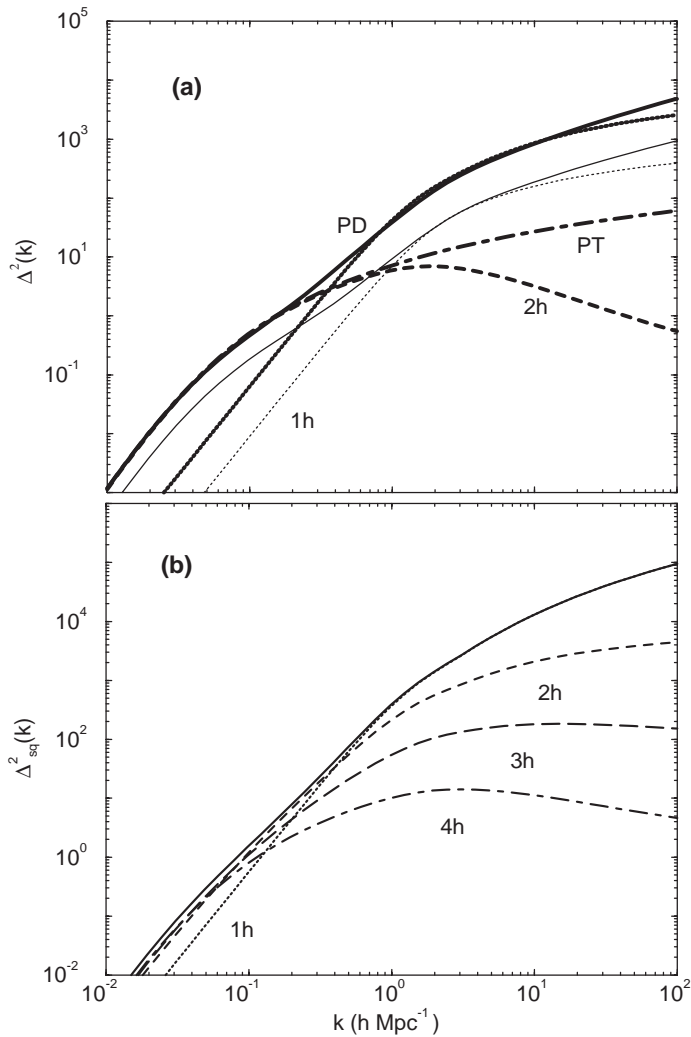,width=3.6in,angle=0}}
\caption{The dark matter power spectrum (a) and square-configuration 
trispectrum (b)
broken into individual contributions under the halo description. 
The lines labeled 'PD' shows the dark matter power spectrum under the
\cite{PeaDod96} (1996) non-linear fitting function while the curve
labeled 'PT' is the linear dark matter power spectrum (at redshift of 0). 
In (a), we show the power spectrum at redshifts of 0 and 1. In (b), we show 
the square configuration trispectrum (see text).
In both (a) and (b), at small scales
the single halo term dominates while at large scales halo 
correlations contribute. }
\label{fig:dmpower}
\end{figure}

Here, we present a semi-analytical estimate of the Fourier
analogue of the four point function, i.e. the trispectrum,
and calculate in detail the configurations that contribute
to power spectrum covariance.
Since weak lensing shear and convergence can be
written as a simple projection of the dark matter density field,
the problem reduces to a study of the trispectrum of
the density field.
Previous studies of the dark matter trispectrum have employed a mix
of perturbation theory and non-linear scalings
(e.g., \cite{Scoetal99} 1999) or N-body simulations (\cite{MeiWhi99} 1999). The former are not applicable
to the full range of scales and configurations of interest;
the latter are limited by computational expense
to a handful of realizations of cosmological models with modest dynamical
range.   

Here, we use the dark-matter halo approach to model the
density field (\cite{Sel00} 2000; 
\cite{MaFry00a} 2000a; \cite{Scoetal00} 2000) and
extend our previous treatments of the two-point and 
three-point lensing statistics (\cite{Cooetal00} 2000; \cite{CooHu00} 2000).
The critical ingredients are: a mass function for the halo
distribution, such as the Press-Schechter (PS; \cite{PreSch74} 1974) 
mass function; a profile
for the dark matter halo, e.g., the profile of \cite{Navetal96} (1996;
NFW),  and a description of halo biasing (\cite{Moetal97} 1997).
In the mildly non-linear regime, where
most of the contribution to lensing is expected, the accuracy of the
halo model has been extensively tested against simulations
at the two point and three point levels
(\cite{Sel00} 2000; \cite{MaFry00a} 2000a; \cite{Scoetal00} 2000).
We present tests here of the four point configurations involved
in the power spectrum covariance.  These techniques can also
be extended to the covariance of the power spectrum of galaxy redshift
surveys with a prescription for assigning galaxies to halos
(\cite{Sel00} 2000; \cite{Scoetal00} 2000). The effect of
non-Gaussianities on the measured galaxy power spectrum, through a
measurement of the angular correlation function, is discussed in
\cite{EisZal99} (1999).

In \S \ref{sec:dm}, we study the trispectrum of the dark matter density
field under the halo model and test it against simulations from
(\cite{MeiWhi99} 1999). 
In \S \ref{sec:covariance}, we apply these techniques to the the
weak lensing covariance and test them against the simulations of
(\cite{WhiHu99} 1999).
We also discuss the effect of power spectrum covariance on cosmological
parameter estimation.

\section{Dark Matter Power Spectrum Covariance}
\label{sec:dm}

We begin by defining the power spectrum, trispectrum and
power spectrum covariance in \S \ref{sec:general}.  We then
derive the halo model for these quantities in \S \ref{sec:halo}.
In \S \ref{sec:dmresults}, we present results and comparisons
with $N$-body simulations.

\subsection{General Definitions} 
\label{sec:general}

The two and four point
correlations of the density field are defined in the usual way
\begin{eqnarray}
\left< \delta(\veck_1) \delta(\veck_2) \right>& = &(2\pi)^3
\delta_\dirac (\veck_{12}) P(k_1) \, , \\
\left< \delta(\veck_1) \ldots \delta(\veck_4)\right>_c &=&
(2\pi)^3 \delta_\dirac (\veck_{1234})  
T(\veck_1,\veck_2,\veck_3,\veck_4) \, ,
\end{eqnarray}
where $\veck_{i\ldots j} = \veck_i + \ldots + \veck_j$ and $\delta_\dirac$ is
the delta function not to be confused with the density perturbation.  Note that
the subscript $c$ denotes the connected piece, i.e. the
trispectrum is defined to be identically zero for a Gaussian field.
Here and throughout, we occasionally suppress the redshift dependence
where no confusion will arise.

Because of the closure condition expressed by the delta function,
the trispectrum may be viewed as a four-sided figure with sides $\veck_i$.  
It can alternately be described by the length of the four sides $k_i$
plus the diagonals.  We occasionally refer to elements of the trispectrum
that differ by the length of the diagonals as different configurations
of the trispectrum.
 
Following \cite{Scoetal99} (1999), we can relate the trispectrum
to the variance of the estimator of the binned power spectrum
\begin{equation}
\hat P_i = {1 \over V } \int_{\shell i} {d^3 k \over V_{\shell i}}
\delta^*(-\veck) \delta(\veck)  \, ,
\end{equation}
where the integral is over a shell in $k$-space centered around $k_i$, 
$V_{\shell i} \approx 4\pi k_i^2 \delta k$ is the volume of the shell and $V$
is the volume of the survey.  Recalling that $\delta({\bf 0}) \rightarrow V/(2\pi)^3$
for a finite volume,
\begin{eqnarray}
C_{ij} &\equiv& \left< \hat P_i \hat P_j \right> - 
	\left< \hat P_i \right> 
	\left< \hat P_j \right>  \nonumber\\
       &=& {1 \over V} \left[ {(2\pi)^3 \over V_{\shell i} } 2 P_i^2 \delta_{ij}+
	T_{ij} \right]  \, ,
\end{eqnarray}
where
\begin{eqnarray}
T_{ij} &\equiv& \int_{\shell i} {d^3 k_i \over V_{\shell i}}
	\int_{\shell j} {d^3 k_j \over V_{\shell j}}
	T(\veck_i,-\veck_i,\veck_j,-\veck_j) \,.
\label{eqn:covarianceij}
\end{eqnarray}
Notice that though both terms
scale in the same way with the volume of the survey, only the Gaussian piece
necessarily decreases with the volume of the shell.  For the Gaussian piece,
the sampling error reduces to a simple root-N mode counting of independent modes
in a shell.  The trispectrum quantifies the non-independence of the modes both within a shell
and between shells.  Calculating the covariance matrix of the power spectrum
estimates reduces to averaging the elements of the trispectrum across configurations
in the shell.  It is to the subject of modeling the trispectrum that we now turn.

\subsection{Halo Model}
\label{sec:halo}

We model the power spectrum and trispectrum of the dark matter
field under the halo approach.  
Here we present in detail the extensions required to model the trispectrum.
We refer the reader to \cite{CooHu00} (2000) for a more in depth treatment of the
ingredients.  

The halo approach models the fully non-linear
dark matter density field as a set of
correlated discrete objects (``halos'') with profiles $\rho_{i}$ 
that for definiteness depend on their mass $M$ and concentration
$c$ as in the NFW profile (\cite{Navetal96} 1996)\footnote{This
prescription can be generalized for more complicated halo profiles in
the obvious way.}
\begin{equation}
\rho(\bfx) = \sum_{i} \rho_\halo (\bfx -  \bfx_i;M_i,c_i) \, ,
\end{equation}
and so a density fluctuation in Fourier space
\begin{eqnarray}
\delta(\veck)& = & \sum_i e^{i \veck \cdot \bfx_i} \delta_{\halo}(\veck;M;c) \\
	     & = & \sum_{V_1,M_1,c_1} 
			n_1 e^{i \veck \cdot \bfx_1} \delta_\halo(\veck,M_1, c_1)\,,
\label{eqn:fourierdelta}
\end{eqnarray}
where we have divided space up into volumes $\delta V$ sufficiently
small such that they contain only one halo $n_1=n_1^2=n_1^\mu = $ 
$1$ or $0$ following \cite{Pee80} (1980).  The final ingredient is that
the halos themselves are taken to be biased tracers of the
{\it linear} density field (denoted PT) such that their number density
fluctuates as
\begin{equation}
{d^2 n \over d M d c}(\bfx) = 
{d^2 \bar n \over d M d c} 
	[b_0 + b_1(M) \delta_\lin(\bfx) + {1 \over 2} b_2(M) \delta_\lin^2(\bf x)
	\ldots]
\end{equation}
where $b_0 \equiv 1$ and
the halo bias parameters are given in \cite{Moetal97} (1997).
Thus 
\begin{eqnarray}
\left< n_1 \right> &=& { d^2 \bar n \over dM dc} \delta M_1 \delta c_1
\, ,\\
\left< n_1 n_2 \right> &=& 
	\left<n_1\right> \delta_{12} + 
	\left<n_1\right> \left<n_2 \right> 
	[b_0^2 + b_1(M_1)b_1(M_2) \, ,
	\nonumber\\
	&& 
		\times \left< \delta_\lin(\bfx_1)\delta_\lin
        (\bfx_2) \right>] \, . \nonumber\\
\left< n_1 n_2 n_3 \right> &=& \ldots \, .
\label{eqn:expectation}
\end{eqnarray}
The derivation of the higher point functions in Fourier space is
now a straightforward but tedious exercise in algebra.  The Fourier transforms
inherent in eqn.~(\ref{eqn:fourierdelta}) convert the correlation
functions in eqn.~(\ref{eqn:expectation}) into the power spectrum, bispectrum,
trispectrum, etc., of perturbation theory.  

Replacing sums
with integrals, we obtain expressions based on the general
integral 
\begin{eqnarray}
I_\mu^\beta(k_1,\ldots,k_\mu) &\equiv&
\int dM \int dc \frac{d^2\bar n}{dM dc} b_\beta(M)  \nonumber\\
&& \quad \delta_\halo(k_1,M,c)\ldots \delta_\halo(k_\mu,M,c)\,.
\label{eqn:integral}
\end{eqnarray}
The index $\mu$ represents the number of points
taken to be in the same halo such that $\left< n_1^\mu \right> =
\left< n_1 \right>$. 

The power spectrum under the halo model becomes (\cite{Sel00} 2000)
\begin{eqnarray}
P(k) &=& P^{1h}(k) +  P^{2h}(k) \,, \\
P^{1h}(k) & = & I_2^0(k,k) \,, \\ 
P^{2h}(k) & = &\left[  I_1^1(k) \right]^2 P^\lin(k)\,,
\end{eqnarray}
where the two terms represent contributions from two points in 
a single halo (1h) and points in different halos (2h) respectively.

Likewise for the trispectrum, the contributions may be separated
into those involving one to four halos
\begin{equation}
T = T^{1h} +  T^{2h} + T^{3h} + T^{4h}\,,
\end{equation} 
where here and below the argument of the trispectrum is understood
to be $(\veck_1,\veck_2,\veck_3,\veck_4)$.
The term involving a single halo probes correlations of dark matter
within that halo
\begin{equation}
T^{1h} =
I_4^0(k_1,k_2,k_3,k_4) \, ,
\end{equation}
and is independent of configuration due to the assumed
spherical symmetry for our halos.

The term involving two halos can be further broken up into two parts 
\begin{equation}
T^{2h} = T^{2h}_{31} + T^{2h}_{22}\,,
\end{equation}
which represent taking three or two points in the first halo
\begin{eqnarray}
T^{2h}_{31} = P^\lin(k_1)I_3^1(k_2,k_3,k_4)I_1^1(k_1) + 3\; {\rm Perm.,} \\
T^{2h}_{22} = P^\lin(k_{12})I_2^1(k_1,k_2)I_2^1(k_3,k_4)+ 2\; {\rm 
Perm.}
\end{eqnarray}
The permutations involve the 3 other choices of $k_i$ for the $I_1^1$ term in
the first equation and the two other pairings of the $k_i$'s for the $I_2^1$ terms
in the second.
Here, we have defined $\veck_{12} =
\veck_1+\veck_2$; note that $k_{12}$ is the length of one of the diagonals
in the configuration.

\begin{figure}[t]
\centerline{\psfig{file=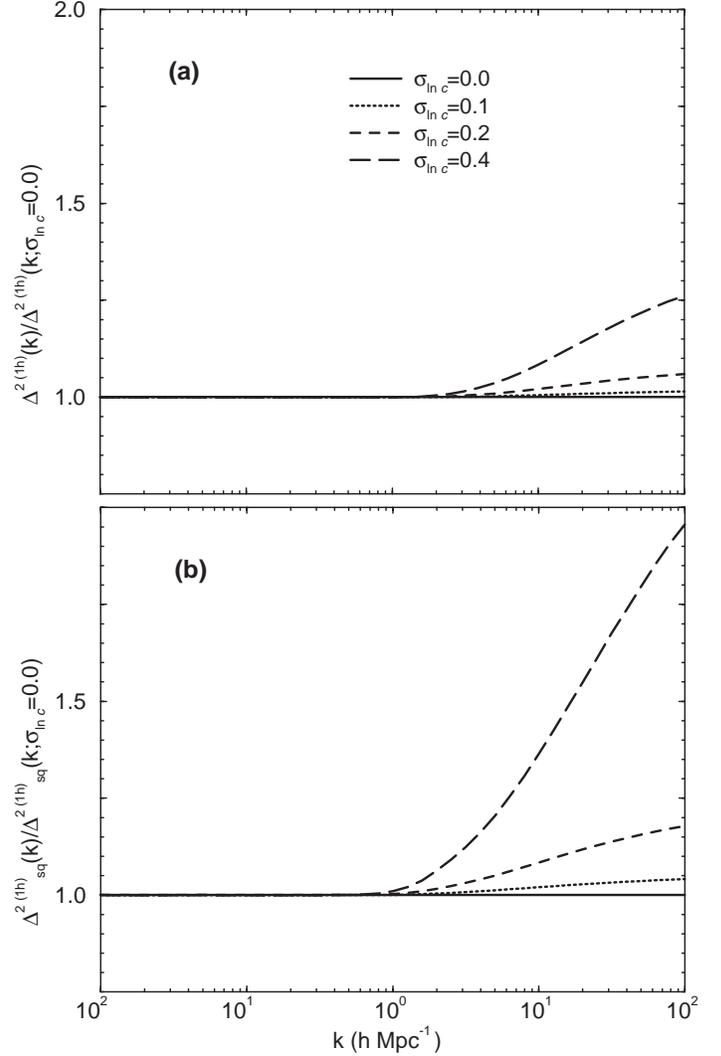,width=3.6in,angle=0}}
\caption{The ratio of the single halo term contribution to that for 
a concentration width $\sigma_{\ln c}\rightarrow 0$ for the
(a) power spectrum and (b) trispectrum. 
The small scale behavior is increasingly sensitive to the high concentration
tails for the higher order statistics.}
\label{fig:conc}
\end{figure}

The term containing three halos can only arise with two points in one halo
and one in each of the others
\begin{eqnarray}
T^{3h} &=& B^\lin(\veck_1,\veck_2,\veck_{34})I_2^1(k_3,k_4)I_1^1(k_1)I_1^1(k_2)
\\
&&+ P^\lin(k_1)P^\lin(k_2)I_2^2(k_3,k_4)I_1^1(k_1)I_1^1(k_2) + 5\; {\rm Perm.}\, ,
\nonumber  
\end{eqnarray}
where the permutations represent the unique pairings of the $k_i$'s in the
$I_2$ factors.  This term also depends on the configuration.
The bispectrum in perturbation theory is given by\footnote{The
kernels $F_n^{\rm s}$ are derived in \cite{Goretal86} (1986) (see,
equations A2 and A3 of \cite{Goretal86} 1986; note that their
$P_n\equiv F_n$), and we have written such that the symmetric form of
$F_n$'s are used. The use of the symmetric form accounts for the
factor of 2 in Eqs.~(\ref{eqn:bpt}) and factors of 4 and 6 in
(\ref{eqn:tript}).}
\begin{eqnarray}
B^\lin(\veck_p,\veck_q,\veck_r) &=& 2 F_2^{\rm s}(\veck_p,\veck_q)P(k_p)P(k_q)
+ 2\; {\rm
Perm.} \, ,
\label{eqn:bpt}
\end{eqnarray}
with $F_2^{\rm s}$ term given by second order gravitational perturbation
calculations (see, below). 

Finally for four halos, the contribution is
\begin{eqnarray}
T^{4h} &=&  I_1^1(k_1)I_1^1(k_2)I_1^1(k_3)I_1^1(k_4) \Big\{ T^\lin 
	+ \Big[ {I_2^1(k_4) \over I_1^1(k_4) }
\nonumber\\ &&\quad \times
P^\lin(k_1)P^\lin(k_2) P^\lin(k_3)+ 3\; {\rm Perm.}\Big] \Big\}, 
\end{eqnarray}
where the permutations represent the choice of $k_i$ in the $I^1$'s in the brackets.
The perturbation trispectrum can 
be written as
(\cite{Fry84} 1984) 
\begin{eqnarray}
&& T^\lin =\nonumber\\
&& \quad 4\left[F_2^{\rm s}(\veck_{12},-\veck_1) F_2^{\rm s}(\veck_{12},\veck_3)P(k_1)P(k_{12})P(k_3)
+ {\rm Perm.}\right] \nonumber \\ 
&&\quad + 6 \left[F_3^{\rm s}(\veck_1,\veck_2,\veck_3)P(k_1)P(k_2)P(k_3) + {\rm 
Perm.}\right]
\, .
\label{eqn:tript}
\end{eqnarray}
The permutations involve a total
of 12 terms in the first set and 4 terms in the second set. 
We now discuss the results from this modeling for a specific choice of
halo input parameters and cosmology.

\begin{table*}
\begin{center}
\caption{\label{tab:dmcorr}}
{\sc Dark Matter Power Spectrum Correlations\\}
\begin{tabular}{lrrrrrrrrrrrr}
\tablevspace{4pt}
\hline
$k$   & 0.031 & 0.044 &  0.058 & 0.074 & 0.093 & 0.110 & 0.138 & 0.169 & 0.206 & 0.254 & 0.313 & 0.385 \\
\hline
0.031 &  1.000 & 0.019 & 0.041 & 0.065 & 0.086 & 0.113 & 0.149 & 0.172 & 0.186&  0.186 & 0.172 & 0.155 \\
0.044 & (-0.017) & 1.000 & 0.036 & 0.075 & 0.111 & 0.153 & 0.204 & 0.238 & 0.261 & 0.264 & 0.251 & 0.230 \\
0.058 &  (0.023) & (0.001) & 1.000 & 0.062 & 0.118 & 0.183 & 0.255 & 0.302 & 0.334 & 0.341 & 0.328 & 0.305 \\
0.074 &  (0.024) & (0.024) & (0.041) & 1.000 & 0.102 & 0.189 & 0.299 & 0.368 & 0.412 & 0.425 & 0.412 & 0.389 \\
0.093 &  (0.042) & (0.056) & (0.027) & (0.079) & 1.000 & 0.160 & 0.295 & 0.404 & 0.466 & 0.485 & 0.475 & 0.453 \\
0.110 &  (0.154) & (0.076) & (0.086) & (0.094) & (0.028) & 1.000 & 0.277 & 0.433 & 0.541 & 0.576 & 0.570 & 0.549 \\
0.138 &  (0.176) & (0.118) & (0.149) & (0.202) & (0.085) & (0.205) & 1.000 & 0.434 & 0.580 & 0.693 & 0.698 & 0.680 \\
0.169 &  (0.188) & (0.180) & (0.138) & (0.229) & (0.177) & (0.251) & (0.281) & 1.000 & 0.592 & 0.737 & 0.778 & 0.766 \\
0.206 &  (0.224) & (0.165) & (0.177) & (0.322) & (0.193) & (0.314) & (0.396) & (0.484) & 1.000 & 0.748 & 0.839 & 0.848 \\
0.254 &  (0.264) & (0.228) & (0.206) & (0.343) & (0.261) & (0.355) & (0.488) & (0.606) & (0.654) & 1.000 & 0.858 & 0.896 \\
0.313 &  (0.265) & (0.234) & (0.202) & (0.374) & (0.259) & (0.397) & (0.506) & (0.618) & (0.720) & (0.816) & 1.000 & 0.914 \\
0.385 &  (0.270) & (0.227) & (0.205) & (0.391) & (0.262) & (0.374) & (0.508) & (0.633) & (0.733) & (0.835) & (0.902) & 1.000 \\
\hline
$\sqrt{\frac{C_{ii}}{C_{ii}^{G}}}$ & 1.00 & 1.01&
1.02& 1.03& 1.04& 1.07& 1.14& 1.23& 1.38& 1.61& 1.90& 2.26 \\
\end{tabular}
\end{center}
\footnotesize
NOTES.---%
Diagonal normalized covariance matrix of the binned dark matter
density field power spectrum with $k$ values in units of h Mpc$^{-1}$.
Upper triangle displays the covariance found under the halo model.
Lower triangle (parenthetical numbers) displays the covariance found
in numerical simulations by \cite{MeiWhi99} (1999).  Final line shows the
fractional increase in the errors (root diagonal covariance) due to non-Gaussianity
as calculated under the halo model.
\end{table*}

\begin{figure}[t]
\centerline{\psfig{file=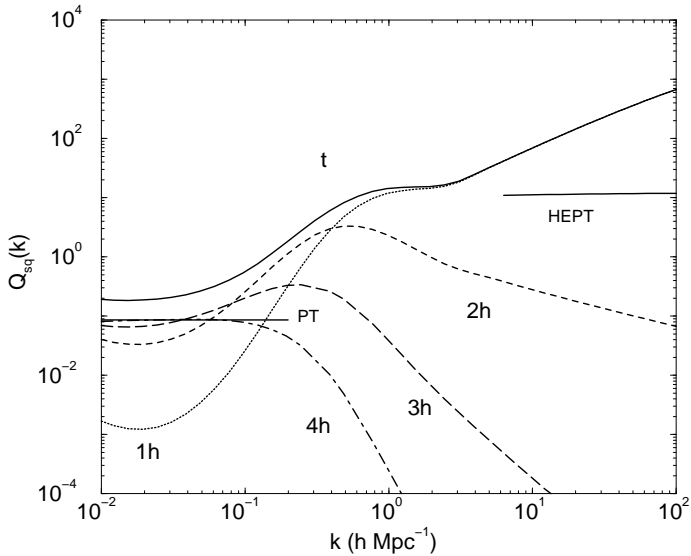,width=3.6in,angle=-90}}
\caption{$Q_{\rm sq}$ at present broken into individual
contributions under the halo description. The hierarchical model predicts
a constant value for $Q_{\rm sq}$ in the deeply
non-linear regime for clustering (HEPT).  In the linear regime,
the perturbation theory (PT) prediction
is reproduced by the $4$ halo term which is only $\sim 1/2$ of the total.  
See text for a discussion of discrepancies.}
\label{fig:q4}

\end{figure}
\subsection{Results}

\label{sec:dmresults}

\subsubsection{Fiducial Model}

We evaluate the trispectrum under the halo model of the last section assuming
an NFW profile for the halos (\cite{Navetal96} 1996) which depends on their
virial mass $M$ and concentration $c$.   For the differential number density
we take
\begin{eqnarray}
{d \bar n \over dM dc} &=& \left({d n \over dM}\right)_{\rm PS} p(c)\,
, \\
p(c) dc &=& \frac{1}{\sqrt{2 \pi \sigma_c^2}} \exp\left[-\frac{(\ln c - 
\ln \bar{c})^2}{2\sigma_{\ln c}^2}\right] d\ln c \, ,\nonumber
\end{eqnarray}
where PS denotes the Press-Schechter mass function.
From the simulations of 
\cite{Buletal00} (2000), the
mean and width of the concentration distribution is
taken to be 
\begin{eqnarray}
\bar{c}(M,z)   & = &  9 (1+z)^{-1} \left[ \frac{M}{M_*(z)}\right]^{-0.13}\,,\\
\sigma_{\ln c} & = &  0.2\,,
\label{eqn:concentration}
\end{eqnarray}
where 
$M_*(z)$ is the non-linear mass scale at which the peak-height
threshold, $\nu(M,z)=1$.

This prescription differs from that in \cite{CooHu00} (2000) where
$\sigma_{\ln c} \rightarrow 0 $ since a finite distribution 
becomes increasingly important for the higher moments.  To maintain
consistency we have also taken the mean concentration directly from
simulations rather than empirically adjust it to match the power 
spectrum.
For the same reason we choose 
a $\Lambda$CDM cosmological model with
$\Omega_m=0.3$, $\Omega_\Lambda=0.7$, $h=0.65$ and a scale invariant
spectrum of primordial fluctuations. This model has
mass fluctuations on the 8 h Mpc$^{-1}$ scale of $\sim$ 1.0,
consistent with the abundance of galaxy clusters (\cite{ViaLid99}
1999) and COBE (\cite{BunWhi97} 1997). For the linear power spectrum,
we take the fitting formula for the transfer function given in
\cite{EisHu99} (1999).

\begin{figure}[t]
\centerline{\psfig{file=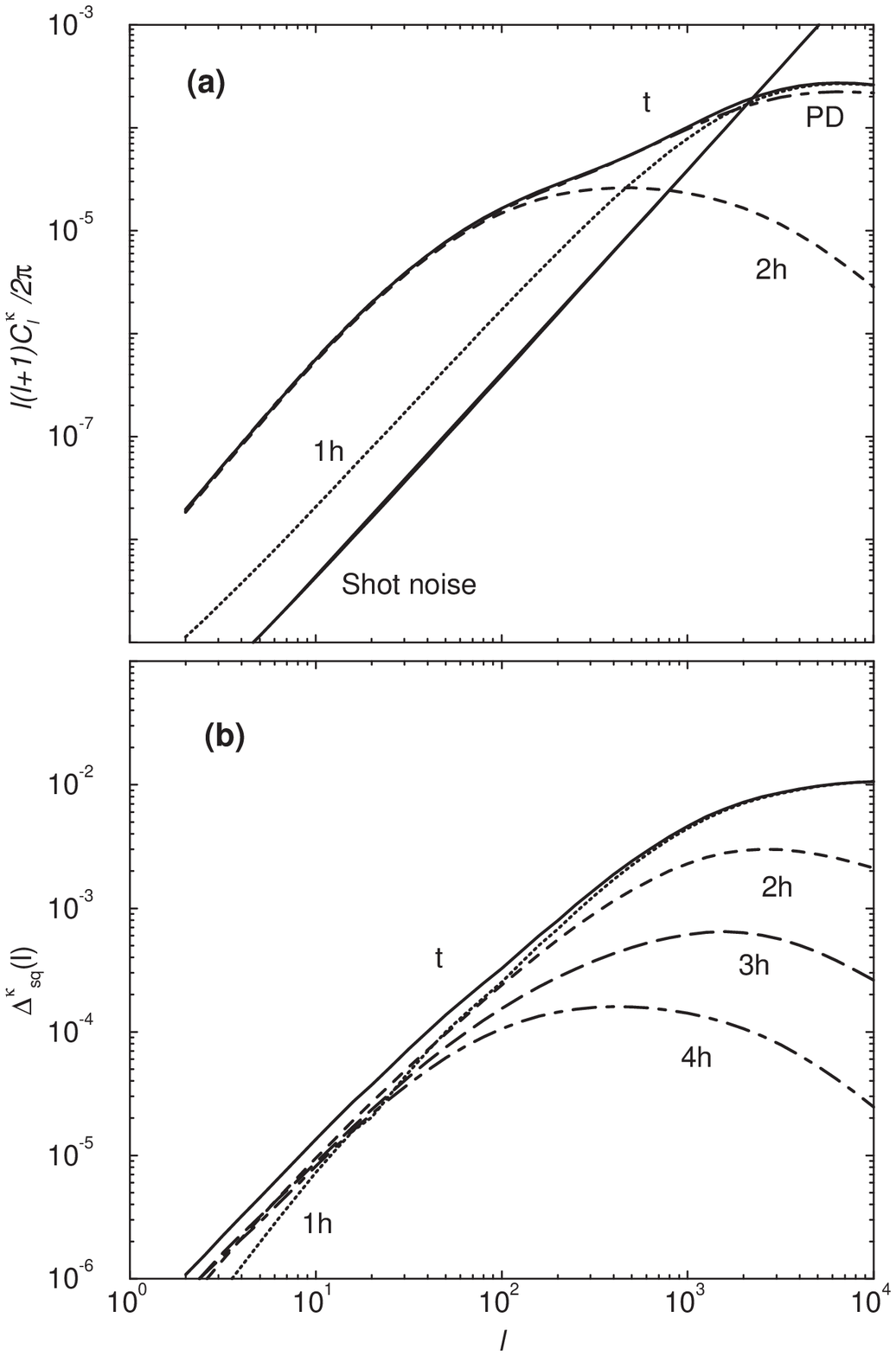,width=3.6in,angle=0}}
\caption{Weak lensing convergence (a) power spectrum and (b)
trispectrum under the halo description. Also shown in (a) is the
prediction from the PD nonlinear power spectrum fitting function. We
have separated individual contributions under the halo approach and
have assumed that all sources are at $z_s=1$. We have also shown the
shot noise contribution to the power spectrum assuming a survey down
to a limiting magnitude of R $\sim$ 25 with an intrinsic rms shear of 0.4 in
each component.}
\label{fig:weakpower}
\end{figure}

\subsubsection{Comparisons}

In Fig.~\ref{fig:dmpower}(a), we show the logarithmic 
power spectrum $\Delta^2(k)=k^3 P(k)/2\pi^2$ with
contributions broken down to the $1h$ and $2h$ terms today and the
$1h$ term at redshift of 1.
We find that there is an
slight overprediction of power at scales  corresponding to
$1 \lesssim k \lesssim 10$ h Mpc$^{-1}$ when compared to the 
\cite{PeaDod96} (1996)
fitting function shown for redshifts of 0 and 1, 
and a more substantial underprediction at small scales
with $k \gtrsim 10$ h Mpc$^{-1}$. Since the non-linear
power spectrum has only been properly studied out to
overdensities $\Delta^2 \sim 10^3$ with numerical simulations
it is unclear whether the small-scale disagreement is significant. 
Fortunately, it is on sufficiently small scales so as not to
affect the lensing observables.

For the trispectrum, we are mainly interested in terms
involving $T(\veck_1,-\veck_1,\veck_2,-\veck_2)$, i.e. parallelograms 
which are defined by either the length $k_{12}$ or the angle
between $\veck_1$ and $\veck_2$.  For illustration purposes
we will take $k_1=k_2$ and the angle to be $90^\circ$ ($\veck_2=\veck_\perp$)
such that
the parallelogram is a square.
It is then convenient to define
\begin{equation}
\Delta^2_{\rm sq}(k) \equiv \frac{k^3}{2\pi^2}
T^{1/3}(\veck,-\veck,\veck_\perp,-\veck_\perp) \, ,
\end{equation}
such that this quantity scales roughly as the logarithmic
power spectrum itself $\Delta^2(k)$.  This spectrum is
shown in Fig.~\ref{fig:dmpower}(b) with the individual
contributions from the 1h, 2h, 3h, 4h terms shown. 
We test the sensitivity
of our calculations to the width of the distribution in Fig.~\ref{fig:conc}, where we
show the ratio between single halo contribution, 
as a function of the
concentration  distribution width, to the halo term with a delta
function distribution $\sigma_c=0$.  
As in the power spectrum
the effect of increasing the width is to increase the amplitude at
small scales due to the high concentration tail of the distribution.
Notice that the width effect
is stronger in the trispectrum than the power spectrum since the
tails of the distribution are weighted more heavily in higher point
statistics. 

To compare the specific scaling predicted by perturbation theory
in the linear regime 
and the hierarchical ansatz in the deeply non-linear regime,
it is useful to define the quantity
\begin{equation}
Q_{\rm sq}(k) \equiv
\frac{T(\veck,-\veck,\veck_\perp,-\veck_\perp)}{[8P^2(k)P(\sqrt{2}k)][4P^3(k)]} \, .
\end{equation}
In the halo prescription, $Q_{\rm sq}$
at $k \simgt 10 k_{\rm nonlin} \sim 10 h$Mpc$^{-1}$
arises mainly from the single halo term. 
In perturbation theory $Q_{\rm sq} \approx 0.085 $. 
The $Q_{\rm sq}$ does not approach the perturbation theory
prediction as $k \rightarrow 0$ since that contribution appears
only as one term in the 4 halo piece.  There is an intrinsic
shot noise error introduced by modeling the continuous density
field by discrete objects.  This error appears 
large in the $Q_{\rm sq}$ statistic since
we have subtracted out the
much larger connected (Gaussian) piece of the four point function.  
For example in the power spectrum covariance,
the error induced by this approximation is much less than the Gaussian
variance. 

The hierarchical ansatz
predicts that $Q_{\rm sq}=$ const. in the deeply non-linear regime.
Its value is unspecified by the ansatz but is given 
as 
\begin{equation}
Q_{\rm sq}^{\rm sat} = \frac{1}{2}\left[\frac{54 - 27\cdot 2^n + 2\cdot 3^n + 6^n}{1+6\cdot2^n + 3\cdot 3^n + 6\cdot 6^n}\right]
\end{equation}
under hyperextended perturbation theory (HEPT; \cite{ScoFri99}).
Here $n=n(k)$ is the linear  power spectral index at $k$. As
shown in Fig.~\ref{fig:q4}, the halo model predicts $Q_{\rm sq}$ 
increases at high $k$.  
This behavior, also present at the three point level for the dark
matter density field bispectrum, 
suggests disagreement between the halo approach and hierarchical clustering ansatz
(see, \cite{MaFry00b} 2000b), though numerical simulations do not yet
have enough resolution  to test this disagreement.  Fortunately
the discrepancy is also outside of the regime important for
lensing.

To further test the accuracy of our halo trispectrum, we compare
dark matter correlations predicted by our method to those from
numerical simulations by \cite{MeiWhi99} (1999).  For this purpose, we
calculate the covariance matrix $C_{ij}$ from Eqn.~(\ref{eqn:covarianceij})
with the bins centered at $k_i$ and volume $V_{\shell i} =
4\pi k_i^2 \delta {k_i}$ corresponding to their scheme. 
We also employ the parameters of their $\Lambda$CDM cosmology
and assume that the parameters that defined the halo
concentration properties from our fiducial $\Lambda$CDM model holds
for this cosmological model also. The physical differences between the two
cosmological model are minor, though normalization differences can 
lead to large changes in the correlation coefficients.

In Table \ref{tab:dmcorr}, we compare the predictions 
for the correlation coefficients
\begin{equation}
\hat C_{ij} = {C_{ij} \over \sqrt{C_{ii} C_{jj}}}
\end{equation}
with the simulations.  Agreement in the off diagonal elements
is typically better than $\pm 0.1$, even in the region where
non-Gaussian effects dominate, and the qualitative
features such as the increase in correlations across the
non-linear scale are preserved.

A further test on the accuracy of the halo approach is to consider 
higher order real-space moments such as skewness and kurtosis. In \cite{CooHu00}
(2000), we discussed the weak lensing convergence skewness under the
halo model and found it to be in agreement with numerical predictions
from \cite{WhiHu99} (1999). The fourth
moment of the density field, under certain approximations, was
calculated by \cite{Scoetal99} (1999) using dark matter halos and
was found to be in good agreement with N-body simulations.
Given that density field moments have already been studied by
\cite{Scoetal99}, we no longer consider them here other than to
suggest that the halo model has provided, at least qualitatively,
a consistent description better than any of the perturbation theory arguments.

Even though the dark matter halo formalism provides a physically
motivated means of calculating the statistics of the dark matter
density field, and especially higher order correlations,
there are several limitations of the approach that should be borne in
mind when interpreting results.
The approach assumes all halos to share a parameterized spherically-symmetric
profile. We have attempted to 
include variations in the halo profiles with the addition of a distribution 
function for concentration parameter based on results from  numerical 
simulations. Unlike our previous calculations presented in 
\cite{Cooetal00} (2000) and \cite{CooHu00} (2000), we have not modified 
concentration-mass relation to fit the PD 
non-linear power spectrum, but rather have taken results directly 
from simulations as inputs. Though we have partly accounted for halo profile variations,
the assumption that halos are spherical is likely to
affect detailed results on the configuration dependence of the
trispectrum. Since we are considering a weighted average of
configurations, our tests here are insufficient to establish the
validity of the trispectrum modeling in general.
Further numerical work is required to 
quantify to what extent the present approach reproduces simulation results
for the full trispectrum.

\begin{figure}[b]
\centerline{\psfig{file=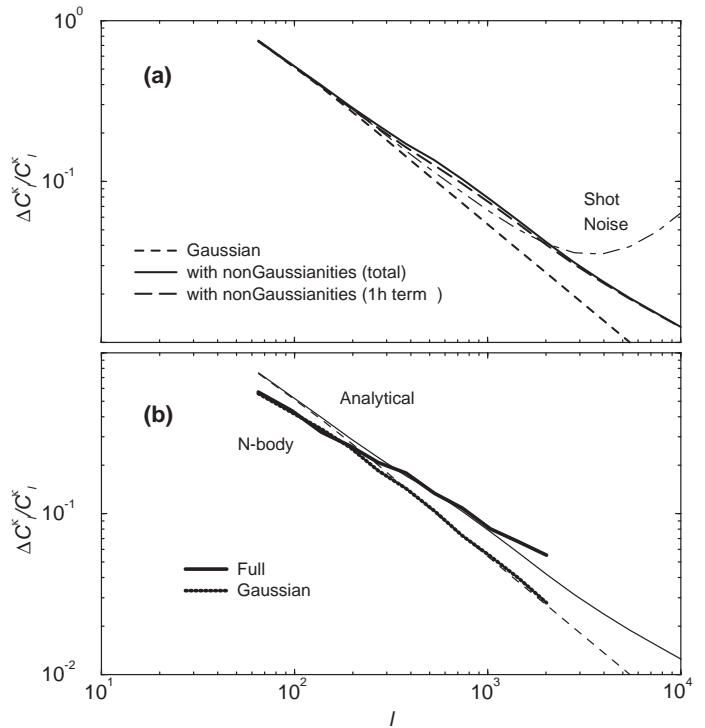,width=3.6in,angle=0}}
\caption{The fractional errors in the measurements of the
convergence band powers.
In (a), we show the fractional errors under the Gaussian approximation,
the full halo description, the Gaussian plus single halo term, and the
Gaussian plus shot noise term (see \S \ref{sec:parameters}). 
As shown, the additional variance can be modeled with the single halo piece
while shot noise generally becomes dominant before non-Gaussian effects
become large.
In (b), we compare the halo model with 
simulations from \cite{WhiHu99} (1999). The decrease in the variance
at small $l$ in the simulations is due to the conversion of variance
to covariance by the finite box size of the simulations.  } 
\label{fig:variance}
\end{figure}

\begin{table*}
\begin{center}
\caption{\label{tab:cov}}
{\sc Weak Lensing Convergence Power Spectrum Correlations\\}
\begin{tabular}{ccccccccccc}
\tablevspace{4pt}
\hline 
$\ell_{\rm bin}$
       & 97      & 138     & 194     & 271     & 378     & 529     & 739 & 1031    & 1440   & 2012 \\
\hline
    97 & 1.00    & 0.04  & 0.05    & 0.07    & 0.08   & 0.09    & 0.09  & 0.09    & 0.08   & 0.08\\
   138 & (0.26) & 1.00   & 0.08    & 0.10   & 0.11    & 0.12    &0.12  & 0.12    & 0.11 & 0.11\\
   194 & (0.12) & (0.31) & 1.00   & 0.14    & 0.17    & 0.18    &0.18 & 0.17    & 0.16 & 0.15\\
   271 & (0.10) & (0.21)  & (0.26) & 1.00  & 0.24   & 0.25     &0.25  & 0.24   & 0.22   & 0.21\\
   378 & (0.02) & (0.09)  & (0.24) & (0.38) & 1.00    & 0.33   &0.33  & 0.32    & 0.30   & 0.28\\
   529 & (0.10) & (0.14)  & (0.28) & (0.33) & (0.45) & 1.00    &0.42  & 0.40    & 0.37  & 0.35\\
   739 & (0.12) & (0.16)  & (0.17)  & (0.34) & (0.38) & (0.50) & 1.00  & 0.48    & 0.45   & 0.42\\
  1031 & (0.15) & (0.18)  & (0.15) & (0.27) & (0.33) & (0.48) & (0.54) & 1.00    & 0.52  & 0.48\\
  1440 & (0.18) & (0.15) & (0.19) & (0.19) &(0.32) & (0.36) & (0.53) & (0.57) & 1.00  & 0.54\\
  2012 & (0.19) & (0.22) & (0.16) & (0.32) & (0.27) & (0.46) & (0.50)
& (0.61) & (0.65) & 1.00\\
\hline
\end{tabular}
\end{center}
\footnotesize
NOTES.---%
Covariance of the binned power spectrum when sources are at a redshift
of 1.
Upper triangle displays the covariance found under the halo model.
Lower triangle (parenthetical numbers) displays the covariance found
in numerical simulations by \cite{WhiHu99} (1999). To be consistent
with these simulations, we use the same binning scheme as the one used there.
\end{table*}

\begin{figure}[t]
\centerline{\psfig{file=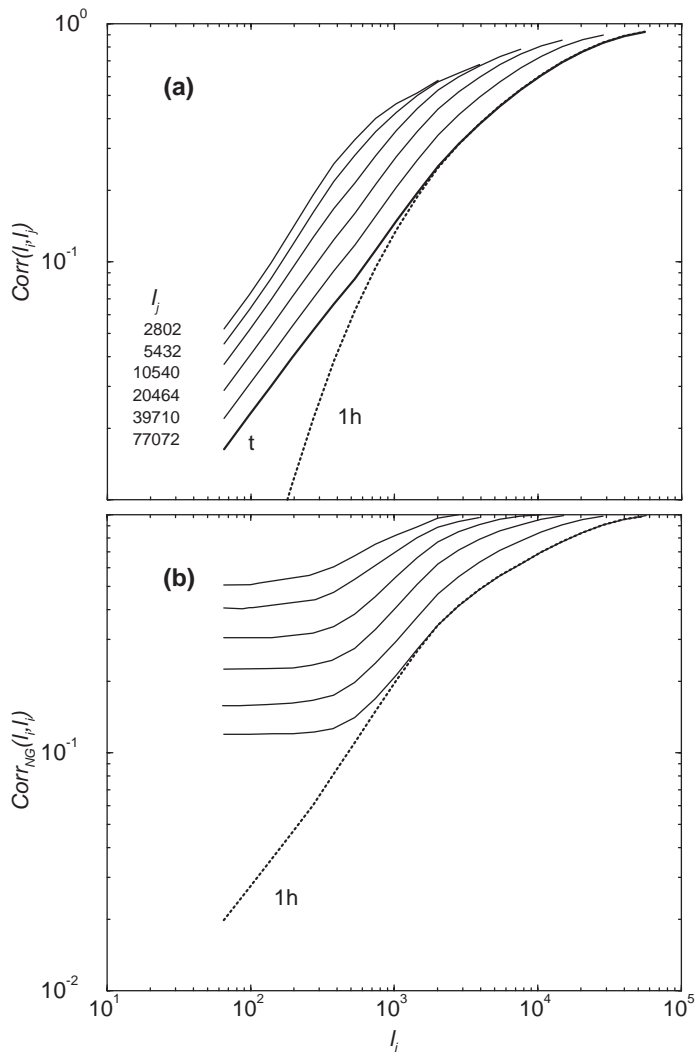,width=3.6in,angle=0}}
\caption{(a) The correlation coefficient, $\hat C_{ij}$ as a function
of the multipole $l_i$ with $l_j$ as shown in the figure.  We show the correlations
calculated with the full halo model and also with only the single halo term for
$l_j=77072$.
In (b), we show
the non-Gaussian correlation coefficient $\hat C_{ij}^{\rm NG}$, 
which only involves the trispectrum (see,
Eq.~\ref{eqn:ng}). The transition to full correlation is due to the 
domination of the single halo contribution. }
\label{fig:corr}
\end{figure}

\section{Convergence Power Spectrum Covariance}
\label{sec:covariance}

\subsection{General Definitions}

Weak lensing probes the statistical properties of the
shear field on the sky which is a weighted projection of
the matter distribution along the line of sight to the
source galaxies.  As such, the observables may be
reexpressed as a scalar quantity, the convergence $\kappa$, on
the sky.  

Its power spectrum and trispectrum are 
defined in the flat sky approximation
in the usual way
\begin{eqnarray}
\left< \kappa(\bfl_1)\kappa(\bfl_2)\right> &=& 
	(2\pi)^2 \delta_\dirac(\bfl_{12}) C_l^\kappa\,,\nonumber\\
\left< \kappa(\bfl_1) \ldots
       \kappa(\bfl_4)\right>_c &=& (2\pi)^2 \delta_\dirac(\bfl_{1234})
	T^\kappa(\bfl_1,\bfl_2,\bfl_3,\bfl_4)\,.
\end{eqnarray}
These are related to the density power spectrum and trispectrum
by the projections (\cite{Kai92} 1992; \cite{Scoetal99} 1999)
\begin{eqnarray}
C^\kappa_l &=& \int d\rad \frac{W(\rad)^2}{d_A^2} P\left
(\frac{l}{d_A};\rad\right) \, , \\
T^\kappa   &=& \int d\rad \frac{W(\rad)^4}{d_A^6} T\left( 
\frac{\bfl_1}{d_A},
\frac{\bfl_2}{d_A},
\frac{\bfl_3}{d_A},
\frac{\bfl_4}{d_A},
;\rad\right) \, ,
\label{eqn:lenspower}
\end{eqnarray}		
where $\rad$ is the comoving distance and  $d_A$ is the angular
diameter distance.  When all background sources are at a distance of
$\rad_s$, the weight function becomes
\begin{equation}
W(\rad) = \frac{3}{2} \Omega_m \frac{H_0^2}{c^2 a} \frac{
d_A(\rad) d_A(\rad_s -\rad)}{d_A(\rad_s)} \, ;
\label{eqn:weight}
\end{equation}
for simplicity, we will assume $\rad_s = r(z_s=1)$. 
In deriving Eq.~(\ref{eqn:lenspower}), we have used the
Limber approximation (\cite{Lim54} 1954) by setting $k=l/d_A$ and
the flat-sky approximation. A potential problem in
using the Limber approximation is that 
we implicitly integrate over the unperturbed photon paths
(Born approximation).   The Born approximation has been tested
in numerical simulations by \cite{JaiSelWhi00} (2000; see
their Fig.~7) and found to be an excellent approximation for
the two point statistics. The same approximation can
also be tested through lens-lens coupling involving lenses
at two different redshifts. For higher order correlations, analytical
calculations in the mildly non-linear regime by 
\cite{vanetal00} (2000b; also, \cite{Beretal97}
1997; \cite{Schetal98} 1998) indicate that corrections are again 
less than a few percent. Thus, our use of the Limber approximation
by ignoring the lens-lens coupling is not expected to 
change the final results significantly.

For the purpose of this calculation, we assume that upcoming weak
lensing convergence power spectrum will measure binned logarithmic band 
powers at several $l_i$'s in multipole space with bins of
thickness $\delta l_i$.
\begin{equation}
\bp_i = 
\int_{\shell i} 
{d^2 l \over{A_{\shell i}}} 
\frac{l^2}{2\pi} \kappa(\bf l) \kappa(-\bf l) \, ,
\end{equation}
where $A_\shell(l_i) = \int d^2 l$ is the area of 2D shell in 
multipole and can be written as $A_\shell(l_i) = 2 \pi l_i \delta l_i 
+ \pi (\delta l_i)^2$.

We can now write the signal covariance matrix
as
\begin{eqnarray}
C_{ij} &=& {1 \over A} \left[ {(2\pi)^2 \over A_{\shell i}} 2 \bp_i^2
+ T^\kappa_{ij}\right]\,,\\
T^\kappa_{ij}&=&
\int {d^2 l_i \over A_{\shell i}} 
\int {d^2 l_j \over A_{\shell j}} {l_i^2 l_j^2 \over (2\pi)^2}
T^\kappa(\bfl_i,-\bfl_i,\bfl_j,-\bfl_j)\,,
\label{eqn:variance}
\end{eqnarray}
where 
$A$ is the area of the survey in steradians.  Again the first
term is the Gaussian contribution to the sample variance and the second
the non-Gaussian contribution.
A realistic survey will also have shot noise variance due to
the finite number of source galaxies in the survey.  
We will return to this point in the \S \ref{sec:parameters}.

\subsection{Comparisons}
\label{sec:discussion}

\begin{table}[b]
\begin{center}
\caption{\label{tab:fisher}}
{\sc Inverse Fisher Matrix ($\times$ 10$^3$)}
\begin{tabular}{lrrrrr}
\tablevspace{4pt}
\hline 
$p_{i}$
	& $\Omega_\Lambda$     & $\ln$ A     & $\Omega_K$     & $n_s$
& $\Omega_mh^2$\\
\hline
$\Omega_\Lambda$ & 1.57  & -5.96  & -1.39    & 4.41 & -1.76\\
$ \ln$ A &  & 25.89   & 5.83    & -17.34 & 6.74\\
$\Omega_K$ & & & 1.41   & -3.81 & 1.43\\
$n_s$ & & & & 14.01 & -6.03\\
$\Omega_mh^2$ & & & & & 2.67\\
\hline
\end{tabular}
\begin{tabular}{lrrrrr}
\tablevspace{3pt}
\hline 
$p_{i}$
	& $\Omega_\Lambda$     & $\ln$ A     & $\Omega_K$     & $n_s$
& $\Omega_mh^2$ \\
\hline
$\Omega_\Lambda$ & 2.03  & -7.84  & -1.82    & 5.76 & -2.30 \\
$ \ln$ A &  & 33.92   & 7.65    & -22.79 & 8.91\\
$\Omega_K$ & & & 1.78   & -5.01 & 1.95 \\
$n_s$ & & & & 18.43 & -7.85\\
$\Omega_mh^2$ & & & & & 3.44 \\
\hline
\end{tabular}
\end{center}

\footnotesize
NOTES.---%
Inverse Fisher matrix under the Gaussian assumption (top) and the
halo model (bottom). The error on an individual parameter is the
square root of the diagonal element of the Fisher matrix for the
parameter while off-diagonal entries of the inverse Fisher matrix
shows correlations, and, thus, degeneracies, between parameters. We
have assumed a full sky survey ($f_\sky=1$) with parameters as
described in \S~\ref{sec:parameters}.
\end{table}

Using the halo model, we can now calculate contributions to lensing
convergence power spectrum and trispectrum. 
The logarithmic power spectrum, shown in Fig.~\ref{fig:weakpower}(a),
shows the same behavior as the density field when compared with
the PD results: a slight overprediction of power when $l \gtrsim 10^3$.
However, these differences are not likely to be observable given
the shot noise from the finite number of galaxies at small scales.

In Fig~\ref{fig:weakpower}(b), we show the scaled trispectrum 
\begin{equation}
\Delta^\kappa_{\rm sq}(l) = \frac{l^2}{2\pi}
T^\kappa(\vecl,-\vecl,\vecl_\perp,-\vecl_\perp)^{1/3} \, .
\end{equation}
where $l_\perp=l$ and $\vecl \cdot \vecl_\perp=0$.
The projected lensing trispectrum again shows the same behavior as the density
field trispectrum with similar conditions on $\veck_i$'s. 

We can now use this trispectrum to study the 
contributions to the covariance, which is what we are primarily
concerned here. In Fig.~\ref{fig:variance}a, we show the
fractional error, 
\begin{equation}
{\Delta \bp_i  \over \bp_i} \equiv {\sqrt{C_{ii}}  \over \bp_i} \, ,
\end{equation}
for bands $l_i$ given in Table~\ref{tab:cov} following the
binning scheme used by \cite{WhiHu99} (1999) on $6^\circ \times 6^\circ$ fields. 
The dashed line compares that with the Gaussian errors, 
involving the
first term in the covariance (Eq.~\ref{eqn:variance}).
At multipoles of a few hundred and
greater, the non-Gaussian term begins to dominate the
contributions.  For this reason, the errors are well approximated by
simply taking the Gaussian and single halo contributions.

In Fig.~\ref{fig:variance}b, we compare these results
with those of the \cite{WhiHu99} (1999) simulations.  The
decrease in errors from the simulations at small $l$ reflects
finite box effects that convert variance to covariance
as the fundamental mode in the box becomes comparable to 
the bandwidth. 

The correlation between the bands
is given by
\begin{equation}
\hat C_{ij} \equiv \frac{C_{ij}}{\sqrt{C_{ii} C_{jj}}} \, .
\end{equation}
In Table \ref{tab:cov} we compare the halo predictions to 
the simulations by \cite{WhiHu99} (1999). 
The upper triangle here is the
correlations under the halo approach, while the lower triangle shows
the correlations found in numerical simulations.
The correlations along individual columns increase (as one goes to
large $l$'s or small angular scales) consistent with simulations.
In Fig.~\ref{fig:corr}, we show the correlation coefficients with (a)
and without (b) the Gaussian contribution to the diagonal. 

We show in Fig.~\ref{fig:corr}(a) the behavior of the correlation
coefficient between a fixed $l_j$ as a function of $l_i$.  When $l_i=l_j$
the coefficient is 1 by definition.  Due to the presence of
the dominant Gaussian contribution at $l_i=l_j$, the coefficient has an apparent 
discontinuity between $l_i=l_j$ and $l_i = l_{j-1}$ that decreases
as $l_j$ increases and non-Gaussian effects dominate.

To better understand this behavior it is useful to isolate
the 
purely non-Gaussian correlation
coefficient 
\begin{equation}
\hat C^{\rm NG}_{ij} =
\frac{T_{ij}}{\sqrt{T_{ii} T_{ij}}} \,.
\label{eqn:ng}
\end{equation}
As shown in Fig.~\ref{fig:corr}(b), 
the coefficient remains constant for $l_i \ll l_j$ and smoothly increases
to unity across a transition scale that is related to where the
single halo terms starts to contribute. 
A
comparison of Fig.~\ref{fig:corr}(b) and \ref{fig:weakpower}(b), shows
that this transition happens around $l$ of few hundred to 1000.
Once the power spectrum is dominated by correlations in single halos,
the fixed profile of the halos will correlate the power in all the modes.
The multiple halo terms on the other hand correlate linear and non-linear
scales but at a level that is generally negligible compared with the
Gaussian variance.

The behavior seen in the halo based covariance, however, is not present when the covariance is
calculated with hierarchical arguments for the trispectrum (see,
\cite{Scoetal99} 1999). With hierarchical arguments, which are by
construction only valid in the deeply nonlinear regime, one predicts
correlations which are, in general, constant across all scales and
shows no decrease in correlations between very small and very large scales.
Such hierarchical models also violate the  Schwarz inequality with
correlations greater than 1 between large and small scales (e.g.,
\cite{Scoetal99} 1999; \cite{Ham00} 2000).
The halo model, however, shows a decrease in correlations similar
to numerical simulations suggesting that the
halo model, at least qualitatively, provides a better
approach to studing non-Gaussian correlations in the translinear regime.

\subsection{Effect on Parameter Estimation}
\label{sec:parameters}

Modeling or measuring the 
covariance matrix of the power spectrum estimates will be
essential for interpreting observational results.
In the absence of many fields where the covariance can be
estimated directly from the data, the halo model provides
a useful, albeit model dependent, quantification of the
covariance.  As a practical approach one could imagine
taking the variances estimated from the survey under
a Gaussian approximation,  but which accounts for uneven 
sampling and edge effects (\cite{HuWhi00} 2000), 
and scaling it up by the non-Gaussian
to Gaussian variance ratio of the halo model along with
inclusion of the band power correlations. Additionally, it is in principle
possible to use the expected correlations 
from the halo model to decorrelate individual band power measurements,
similar to studies involving CMB temperature anisotropy and galaxy
power spectra (e.g., \cite{Ham97} 1997; \cite{HamTeg00} 2000).

We can estimate the resulting effects on cosmological parameter
estimation with an analogous procedure on the Fisher matrix.
In \cite{HuTeg99} (1999), the potential of wide-field lensing
surveys to measure cosmological parameters was investigated
using the Gaussian approximation of a diagonal covariance
and Fisher matrix techniques.
The Fisher matrix is simply a projection of the covariance
matrix onto the basis of cosmological parameters $p_i$
\begin{equation}
{\bf F}_{\alpha\beta} = \sum_{ij} 
 	{\partial \bp_i \over \partial p_\alpha} (C_{\rm tot}^{-1})_{ij}
{\partial \bp_j \over \partial p_\beta} \, ,
\label{eqn:fisher}
\end{equation}
where the total covariance includes both the signal
and noise covariance.  Under the approximation of Gaussian shot
noise, this reduces to replacing $C^\kappa_l \rightarrow
C^\kappa_l + C^{\rm SN}_l$ in the expressions leading up
to the covariance equation~(\ref{eqn:variance}).  
The shot noise power spectrum is given by 
\begin{equation}
C^{\rm SN}_l = \frac{\langle \gamma_{\rm int}^2\rangle}{\bar{n}} \, ,
\end{equation}
where $\langle \gamma_{\rm int} \rangle^{1/2} \sim 0.4$ is the
rms noise per component introduced by intrinsic ellipticities and
measurement errors and $\bar{n} \sim 6.6
\times 10^{8}$ sr$^{-1}$ is the surface number density of background
source galaxies. The numerical values here are appropriate
for surveys that reach a  
limiting magnitude in $R\sim 25$ (e.g., \cite{Smaetal95} 1995).

Under the approximation that there are a sufficient number
of modes in the band powers that the distribution of power
spectrum estimates is approximately Gaussian, the Fisher matrix quantifies
the best possible errors on cosmological parameters that can
be achieved by a given survey.  In particular $F^{-1}$ is
the optimal covariance matrix of the parameters and $(F^{-1})_{ii}^{1/2}$
is the optimal error on the $i$th parameter.
Implicit in this approximation of the Fisher matrix is the neglect of information
from the cosmological parameter dependence of the covariance matrix
of the band powers themselves.  Since the covariance is much less
than the mean power, we expect this information content to be
small. 

In order to estimate the effect of non-Gaussianities on the
cosmological parameters, we calculate the Fisher matrix elements using
our fiducial $\Lambda$CDM cosmological model and define the dark matter
density field, today, as
\begin{equation}
\Delta^2(k) = A^2 \left( \frac{k}{H_0} \right)^{n_s+3} T^2(k) \, .
\end{equation}
Here, $A$ is the amplitude of the present day density fluctuations 
and $n_s$ is the tilt at the Hubble scale. The density power spectrum
is evolved to higher redshifts using the growth function $G(z)$
(\cite{Pee80} 1980) and the transfer function $T(k)$ is calculated
using  the fitting functions from \cite{EisHu99} (1999). Since we are
only interested in the relative effect of non-Gaussianities, 
we restrict ourselves to a small subset of the cosmological parameters
considered by \cite{HuTeg99} (1999) and assume a full sky survey with $f_\sky=1$.

In Table~3, we show the inverse Fisher matrices determined under the
Gaussian and non-Gaussian covariances, respectively. 
For the purpose of this calculation, we adopt the binning scheme as 
shown in Table~2, following \cite{WhiHu99} (1999).
The Gaussian errors are computed using the same scheme by setting
$T^\kappa =0$.  As shown in Table~3, the
inclusion of non-Gaussianities lead to an increase in the inverse
Fisher matrix elements.
We compare the errors on individual parameters, mainly
$(F^{-1})_{ii}^{1/2}$, between the Gaussian and
non-Gaussian assumptions in Table~{\ref{tab:errors}}.  
The errors increase typically by $\sim 15$\%.  
Note also that band power correlations do not necessarily increase
cosmological parameter errors.  Correlations induced by non-linear 
gravity introduce larger errors in the overall amplitude of the power 
spectrum measuremenents but have a much smaller effect on those 
parameters controlling the shape of the spectrum.

For a survey of this assumed depth, the shot noise power becomes
the dominant error before the non-Gaussian signal effects dominate
over the Gaussian ones.   For a deeper survey with better imaging,
such as the one planned with Large-aperture Synoptic Survey Telescope
(LSST; \cite{TysAng00} 2000)\footnote{http://www.dmtelescope.org}, the effect of shot noise
decreases and non-Gaussianity is potentially more
important. However, the non-Gaussianity itself also decreases with
survey depth, and as we now discuss, in terms of the effect of
non-Gaussianities, deeper surveys should be preferred over the shallow ones.

\subsection{Scaling Relations}

To better understand how the non-Gaussian contribution scale with our
assumptions,  we consider the ratio of
non-Gaussian variance to the Gaussian variance (\cite{Scoetal99} 1999), 
\begin{equation}
\frac{C_{ii}}{C_{ii}^{\rm G}} = 1 + R \, ,
\end{equation}
with
\begin{equation}
R \equiv \frac{A_{si} T_{ii}^\kappa}{(2
\pi)^2 2 C_i^2} \, .
\end{equation}
Under the assumption that contributions to lensing convergence can be
written through an effective distance $r_\star$, at half the angular
diameter distance to background sources, and a width $\Delta r$
for the lensing window function,
the ratio of lensing convergence trispectrum and power
spectrum contribution to the variance  can be further simplified to
\begin{equation}
R \sim \frac{A_{si}}{V_{\rm eff}}\frac{ \bar{T}(r_\star)}{\bar{P}^2(r_\star)} \, .
\end{equation}
Since the lensing window function peaks at $r_\star$, we have
replaced the integral over the window function of the
density field trispectrum and power spectrum by its value at the peak.
This ratio shows how the relative contribution from non-Gaussianities
scale with survey parameters: (a) increasing the bin size, through
$A_{si}$ ($\propto \delta l$), leads to an increase in the non-Gaussian contribution linearly,
(b) increasing the source redshift, through the effective volume of
lenses in the survey 
($V_{\rm eff} \sim r_\star^2 \Delta r$), decreases the non-Gaussian contribution, while (c)
the growth of the density field trispectrum and power spectrum,
through the ratio $\bar{T}/\bar{P}^2$,
decreases the contribution as one moves to a higher redshift. The
volume factor quantifies the number of foreground halos in the survey
that effectively act as gravitational lenses 
for background sources; as the number of such halos is increased, 
the non-Gaussianities are reduced by the central limit theorem.

In order to determine whether its the increase in volume or the
decrease in the growth of
structures that lead to a decrease in the relative importance of non-Gaussianities
as one moves to a higher source redshift, we numerically calculated
the non-Gaussian to Gaussian variance ratio under the halo model for several source redshifts and
survey volumes. Up to source redshifts $\sim$ 1.5, the increase in
volume decreases the non-Gaussian contribution significantly. When
surveys are sensitive to sources at redshifts beyond 1.5, 
the increase in volume becomes less significant
and the decrease in the growth of structures
begin to be important in
decreasing the non-Gaussian contribution. Since, in the deeply non-linear
regime, $\bar{T}/\bar{P}^2$ scales with redshift as the cube of the growth factor,
this behavior is consistent with the overall redshift scaling of the
volume and growth. 

Given that scalings
always lead to decrease the effect of non-Gaussianities in deep
lensing surveys, with a decrease in the shot noise contribution also, 
shallow surveys that only probe sources out to 
redshifts of a few tenths are more likely to be dominated by
non-Gaussianities; such shallow surveys are also likely to be affected by 
intrinsic correlations of galaxy shapes (e.g., \cite{Catetal00}
2000; \cite{CroMet00} 2000; \cite{Heaetal00} 2000). These possibilities,
generally, suggests that deeper surveys should be preferred over
shallow ones for weak lensing purposes.

\begin{table}
\begin{center}
\caption{\label{tab:errors}}
{\sc Parameter Errors}
\begin{tabular}{lrrrrr}
\tablevspace{4pt}
\hline 
	& $\Omega_\Lambda$     & $\ln$ A     & $\Omega_k$     & $n_s$
& $\Omega_m h^2$ \\
\hline
Gaussian  & 0.039  & 0.160    & 0.037    & 0.118 & 0.051\\
Full & 0.045  & 0.184   & 0.042    & 0.135 & 0.058\\
Increase (\%) & 15.3 & 15 & 13.5   & 14.4 & 13.7\\
\hline
\end{tabular}
\end{center}

\footnotesize
NOTES.---%
Parameter errors, $(F^{-1})_{ii}^{1/2}$, under the Gaussian assumption (top) and the
halo model (bottom) and following the inverse-Fisher matrices in
Table~3. We have assumed a full sky survey ($f_\sky=1$) with parameters as
described in \S~\ref{sec:parameters}.
\end{table}

\section{Conclusions}
\label{sec:conclusion}  

Weak gravitational lensing due to large scale structure
provides important information on the evolution
of clustering and angular diameter distances and therefore, cosmological parameters.  
This
information complements what can be learned from cosmic microwave
background anisotropy observations. The tremendous progress on the
observational front warrants detailed studies of the statistical properties
of the lensing observables and their use in constraining cosmological
models.

The non-linear growth of large-scale structure induces
high order correlations in the derived shear and convergence fields.
In this work, we have studied the four point correlations in the
fields.  Four point statistics are special in that they quantify 
the errors in the determination of the two point statistics.
To interpret future lensing measurements on the power spectrum, it will
be essential to have an accurate assessment of the correlation 
between the measurements. 

Using the halo model for clustering, we have provided a semi-analytical
method to calculate the four point function of the
lensing convergence as well as the dark matter density field. 
We have tested this model
against numerical $N$-body simulations of the power spectrum covariance
in both the density and
convergence fields and obtained good agreement.  As such, this
method provides a practical means of estimating the error matrix
from future surveys in the absence of sufficiently large fields where
it may be estimated directly from the data or large suites of $N$-body 
simulations where it can be quantified in a given model context.
Eventually a test of whether the covariance matrix estimated from
the data and the theory agree may even provide further cosmological constraints.
 
This method may also be used to study other aspects of the four point
function in lensing and any field whose relation to the dark matter
density field can be modeled.  Given the approximate nature of these
approximations, each potential use must be tested against simulations.
Nonetheless, the halo model provides the most intuitive and
extensible means to study non-Gaussianity in the cosmological context
currently known.

\acknowledgments
We acknowledge 
useful discussions with  Dragan Huterer, 
Roman Scoccimaro, Uros Seljak, Ravi Sheth and
Matias Zaldarriaga. WH is supported by the Keck Foundation.


\begin{thebibliography}{99}
\frenchspacing

\bibitem[Bacon et al]{Bacetal00}
	Bacon, D., Refregier, A., Ellis R. 2000, MNRAS submitted, astro-ph/0003008

\bibitem[Bartelmann \& Schneider]{BarSch00}
	Bartelmann, M., Schnerider, P. 2000, Physics Reports in press, astro-ph/9912508

\bibitem[Bernardeau et al]{Beretal97}
	Bernardeau, F., van Waerbeke, L., Mellier, Y. 1997, A\&A, 322, 1

\bibitem[Blandford et al]{Blaetal91} 
	Blandford, R. D., Saust, A. B., Brainerd, T. G., Villumsen, J. V. 1991, MNRAS 251, 60

\bibitem[Bullock et al]{Buletal00}
	Bullock, J. S., Kolatt, T. S., Sigad, Y. et al. 2000, MNRAS in
press, astro-ph/9908159

\bibitem[Bunn \& White]{BunWhi97}
       Bunn, E. F., White, M. 1997, ApJ, 480, 6

\bibitem[Catelan et al]{Catetal00}
        Catelan, P., Kamionkowski, M., Blandford, R. D. 2000, MNRAS
submitted, astro-ph/0005470


\bibitem[Cooray]{Coo99}
	Cooray, A. R. 1999, A\&A, 348, 673

\bibitem[Cooray \& Hu]{CooHu00}
	Cooray, A., Hu, W. 2000, ApJ in press, astro-ph/0004151

\bibitem[Cooray et al]{Cooetal00}
	Cooray, A., Hu, W., Miralda-Escud\'e, J. 2000b, ApJ 536, L9

\bibitem[Croft \& Metzler]{CroMet00}
        Croft, R. A., \& Metzler, C. 2000, ApJ submittted, astro-ph/0005384

\bibitem[Eisenstein \& Hu]{EisHu99}
        Eisenstein, D.J. \& Hu, W. 1999, ApJ, 511, 5


\bibitem[Eisenstein \& Zaldarriaga]{EisZal99}
	Eisenstein, D. J. \& Zaldarriaga, M. 1999, ApJ in press, astro-ph/9912149

\bibitem[Fry]{Fry84}
Fry, J. N. 1984, ApJ, 279, 499

\bibitem[Goroff et al]{Goretal86}
	Goroff, M. H., Grinstein, B., Rey, S.-J., Wise, M. 1986, ApJ,
311, 6

\bibitem[Hamilton]{Ham97}
	Hamilton, A. J. S. 1997, MNRAS, 289, 285

\bibitem[Hamilton]{Ham00}
	Hamilton, A. J. S. 2000, MNRAS, 312, 257

\bibitem[Hamilton \& Tegmark]{HamTeg00}
	Hamilton, A. J. S. \& Tegmark, M. 2000, MNRAS, 312, 285

\bibitem[Heavens et al.]{Heaetal00}
        Heavens, A., Refregier, A., Heymans, C. 2000, MNRAS submitted,
        astro-ph/0005269

\bibitem[Henry]{Hen00}
        Henry, J. P. 2000, ApJ in press, astro-ph/0002365

\bibitem[Hu \& Tegmark]{HuTeg99} Hu W., Tegmark M. 1999, ApJ 514, L65

\bibitem[Hu \& White]{HuWhi00} Hu W., White M. 2000, ApJ submitted, astro-ph/0010352

\bibitem[Hui]{Hui99} Hui, L. 1999, ApJ, 519, L9

\bibitem[Jain \& Seljak]{JaiSel97}  Jain B., Seljak U. 1997, ApJ 484,
560

\bibitem[Jain et al]{JaiSelWhi00} Jain, B., Seljak, U. \& White, M. 2000, ApJ 530, 547

\bibitem[Jing]{Jin00} Jing, Y. P. 2000, ApJ, 535, 30.

\bibitem[Jing \& Suto]{JinSut00} Jing, Y. P., \& Suto, Y. 2000, ApJ,
529, L69

\bibitem[Kaiser]{Kai92}
      Kaiser, N. 1992, ApJ, 388, 286

\bibitem[Kaiser]{Kai98}
      Kaiser, N. 1998, ApJ, 498, 26

\bibitem[Kaiser et al]{Kaietal00}
      Kaiser, N., Wilson, G., Luppino, G.A., 2000,
	ApJ submitted, astro-ph/0003338

\bibitem[Limber]{Lim54}
        Limber, D. 1954, ApJ, 119, 655

\bibitem[Ma \& Fry]{MaFry00a}
	Ma, C.-P., Fry, J. N. 2000a, ApJ submitted, astro-ph/0003343

\bibitem[Ma \& Fry]{MaFry00b}
	Ma, C.-P., Fry, J. N. 2000b, ApJ, 538, L107

\bibitem[Meiksin \& White]{MeiWhi99}
	Meiksin, A. \& White, M. 1999, MNRAS, 308, 1179

\bibitem[Miralda-Escud\'e]{Mir91} Miralda-Escud\'e J. 1991, ApJ 380, 1

\bibitem[Mo et al.]{Moetal97}
        Mo, H. J., Jing, Y. P., White, S. D. M. 1997, MNRAS, 284, 189

\bibitem[Navarro et al]{Navetal96}
	Navarro, J., Frenk, C., White, S. D. M., 1996, ApJ, 462, 563 [NFW]

\bibitem[Peacock \& Dodds]{PeaDod96}
        Peacock, J.A., Dodds, S.J. 1996, MNRAS, 280, L19

\bibitem[Peebles]{Pee80}
	Peebles, P.J.E. 1980, {\it The Large-Scale Structure of the Universe},
	Princeton University Press, Princeton, NJ.
  
\bibitem[Press \& Schechter]{PreSch74}
	Press, W. H., Schechter, P. 1974, ApJ, 187, 425 [PS]

\bibitem[Scherrer \& Bertschinger]{SchBer91}
	Scherrer, R.J., Bertschinger, E. 1991, ApJ, 381, 349	

\bibitem[Schneider et al]{Schetal98}
        Schneider P., van Waerbeke, L., Jain, B., Guido, K. 1998,
MNRAS, 296, 873

\bibitem[Scoccimarro, Zaldarriaga \& Hui]{Scoetal99}
	Scoccimarro, R., Zaldarriaga, M. \* Hui, L. 1999, ApJ, 527, 1

\bibitem[Scoccimarro et al.]{Scoetal00}
	Scoccimarro, R., Sheth, R., Hui, L. \& Jain, B. 2000, astro-ph/0006319

\bibitem[Scoccimarro \& Frieman]{ScoFri99}
        Scoccimarro, R. \& Frieman, J. 1999, ApJ, 520, 35

\bibitem[Seljak]{Sel00}
	Seljak, U. 2000, PRD submitted, astro-ph/0001493
 


\bibitem[Smail et al]{Smaetal95}
	Smail, I., Hogg, S. W., Yan, L., \& Cohen, J. G. 1995, ApJ,
449, L105

\bibitem[Tyson \& Angel]{TysAng00}
		Tyson, A., Angel, R. 2000, {\it The Large-Aperture
Synoptic Survey Telescope}, in ``New Era of Wide-Field Astronomy'',
ASP Conference Series.

\bibitem[Van Waerbeke et al]{Vanetal99} 
	Van Waerbeke, L., Bernardeau, F., Mellier, Y. 1999, A\&A, 342, 15

\bibitem[Van Waerbeke et al]{Vanetal00} 
	Van Waerbeke, L., Mellier, Y., Erben, T. et al. 2000a, A\&A
submitted, astro-ph/0002500	

\bibitem[Van Waerbeke et al]{vanetal00}
        Van Waerbeke, L., Hamana, T., Scocimarro, R., Colombi, S.,
Bernardeau, F. 2000b, MNRAS submitted, astro-ph/0009426

\bibitem[Viana \& Liddle]{ViaLid99}
        Viana, P. T. P., Liddle, A. R. 1999, MNRAS, 303, 535

\bibitem[White \& Hu]{WhiHu99}
	White, M., Hu, W. 1999, ApJ, 537, 1

\bibitem[Wittman et al]{Witetal00}
	Wittman, D. M., Tyson, J. A., Kirkman, D., Dell'Antonio, I.,
Bernstein, G. 2000, Nature submitted, astro-ph/0003014

 

\end{thebibliography}
\end{document}